\documentclass[11pt]{article}
\usepackage[a4paper,margin=25mm]{geometry}
\usepackage{amsmath,amssymb,amsthm}
\usepackage{lmodern,microtype,tikz}
\usetikzlibrary{arrows.meta,decorations.pathmorphing}
\usepackage[numbers,sort&compress]{natbib}
\usepackage[colorlinks=true,linkcolor=blue!45!black,citecolor=blue!45!black,urlcolor=blue!45!black]{hyperref}
\usepackage[font=small,labelfont=bf]{caption}
\theoremstyle{definition}

\newcommand{\ket}[1]{\lvert #1\rangle}

\newcommand{\vev}[1]{\langle #1\rangle}
\newcommand{\tc}{\tau_c}
\newcommand{\Gc}{G_{\mathrm{cont}}}

\DeclareMathOperator{\Rea}{Re}
\DeclareMathOperator{\Ima}{Im}

\begin{document}
\begin{center}
{\LARGE Fate of Bouncing Singularities at Finite $G_N$\par}
\vspace{8pt}
{\large Zohar Komargodski\par}\vspace{4pt}{\small\itshape Simons Center for Geometry and Physics, Stony Brook University,\\ Stony Brook, NY, USA\par}
\end{center}
\vspace{2pt}
\begin{abstract}
The gravity approximation to a thermal two-point function can develop a complex-time singularity associated with a trajectory that reaches the black-hole singularity. We show that the corresponding singularity is absent at finite $G_N$. This is done by extending Conformal Field Theory analyticity properties to the second sheet (and to the  universal cover). 
We also  put two different upper bounds on the amplitude of the  correlator at the would-be singularity, by certain finite partial sums, and by a certain partition function on a  manifold with nontrivial topology, where two thermal geometries are glued.
We discuss the possible implications for the black hole singularity. 
\end{abstract}

\section{Introduction and Summary}

Complexified trajectories that reach the singularity of an AdS black hole can leave singularities in analytically continued boundary correlators \cite{KOS,FHKS,FL}. These \emph{bouncing singularities} are a feature (or, perhaps,  a bug?) of a semiclassical approximation. Their status in the exact finite-$N$ theory\footnote{By finite $N$ we mean finite $C_T$, and also e.g. that the theory has finite heat capacity per unit volume. This translates to finite $G_N$ in holographic theories.} is the topic of this short note. 

We consider a holographic CFT at temperature $1/\beta$, and we fix a scalar operator $O$ of large dimension $\Delta_O\gg1$. To study its correlation functions, we need to analyze some properties of the dual geometry. 
In the approximation that the bulk is given by the Einstein-Hilbert theory, the thermal state is the planar black brane. In units of the AdS radius,
\begin{equation}
ds^2=-f\,dt^2+\frac{d\varrho^2}{f}+\varrho^2d\mathbf x^2,\qquad
f(\varrho)=\varrho^2\Bigl(1-\frac{\varrho_h^d}{\varrho^d}\Bigr),\qquad
\beta=\frac{4\pi}{d\,\varrho_h},
\label{eq:brane}
\end{equation}
with the boundary at $\varrho=\infty$ and the singularity at $\varrho=0$. The two point function of the operator $O$ is $e^{-\Delta_O L}$, where $L$ is the regularised length of a geodesic joining the two boundary points \cite{KOS,FHKS,FL}.

\begin{figure}[!ht]
\centering
\begin{tikzpicture}[>=Stealth,font=\small,x=2cm,y=2cm]
\fill[blue!5] (-1,-1)--(-1,1) .. controls (-.4,.68) and (.4,.68) .. (1,1)--(1,-1) .. controls (.4,-.68) and (-.4,-.68) .. (-1,-1)--cycle;
\draw[gray] (-1,-1)--(1,1) (-1,1)--(1,-1);
\draw[dashed,gray] (-1,0)--(1,0);
\draw[thick] (-1,-1)--(-1,1) (1,-1)--(1,1);
\draw[thick,decorate,decoration={zigzag,segment length=3pt,amplitude=.8pt}] (-1,1) .. controls (-.4,.68) and (.4,.68) .. (1,1);
\draw[thick,decorate,decoration={zigzag,segment length=3pt,amplitude=.8pt}] (-1,-1) .. controls (-.4,-.68) and (.4,-.68) .. (1,-1);
\draw[red!70!black,densely dotted,thick] (1,-.12) .. controls (.4,.44) and (.12,.64) .. (0,.64) .. controls (-.12,.64) and (-.4,.44) .. (-1,-.12);
\draw[red!70!black,thick] (1,-.24)--(0,.76)--(-1,-.24);
\draw[->,red!70!black,thick] (1,-.24)--(.45,.31);
\draw[->,red!70!black,thick] (0,.76)--(-.55,.21);
\fill[red!70!black] (1,-.24) circle (.025) (-1,-.24) circle (.025);
\node[right] at (1.03,-.24) {$O$};
\node[left] at (-1.03,-.24) {$O$};
\node[gray,fill=blue!5,inner sep=1pt] at (-.5,0) {$t=0$};
\node[gray,rotate=-45,fill=blue!5,inner sep=1pt] at (.52,-.52) {horizon};
\node at (0,1.1) {singularity, $\varrho=0$};
\node[below] at (-1,-1.02) {second boundary};
\node[below] at (1,-1.02) {first boundary};
\end{tikzpicture}
\caption{The maximally extended black brane for $d>2$: boundaries (thick), horizons (grey) and singularities (zigzag). Solid red: the radial null ray that leaves one boundary, reflects off the singularity at its midpoint and reaches the other boundary. The coordinate time that elapses along it is real, and equal to $(\beta/2)\cot(\pi/d)$. Dotted: a spacelike geodesic of large energy, which hugs the ray. }
\label{fig:penrose}
\end{figure}
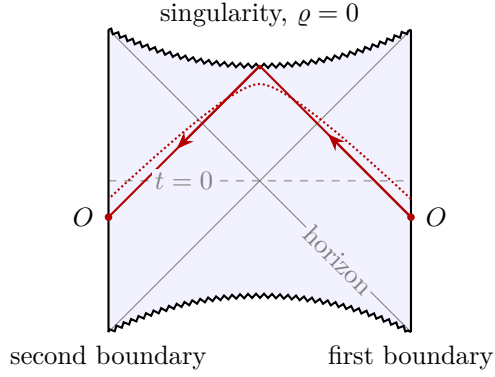

The geometry~\eqref{eq:brane} is two-sided. Its maximal extension  is shown in Figure~\ref{fig:penrose}, where we see two asymptotic boundaries joined through the interior, and describes two copies of the CFT in the thermofield-double state. 
For our purposes the second copy is a bookkeeping device: the two-boundary correlator is the thermal correlator at Lorentzian separation $t-t'-i\beta/2$, where $t,t'$ are Killing times on the first and second boundaries, respectively \cite{KOS,FHKS}.\footnote{Let $O$ be a Hermitian scalar operator at $t=0$, $\mathbf x=0$ in a single CFT with Hamiltonian $H$, and $O(t)=e^{iHt}Oe^{-iHt}$. On the doubled Hilbert space we define $O_1(0)=O\otimes\mathbf 1$ and $O_2(0)=\mathbf 1\otimes(\Theta O\Theta^{-1})$, where $\Theta$ is the antiunitary CPT map to the second copy. The thermofield-double state is $|\mathrm{TFD}\rangle=Z^{-1/2}\sum_n e^{-\beta E_n/2}|n\rangle_1|\bar n\rangle_2$, with $H|n\rangle=E_n|n\rangle$ and $|\bar n\rangle=\Theta|n\rangle$. We use the Killing time, which increases upwards on the first boundary of Figure~\ref{fig:penrose} and downwards on the second. Its generator $H_1-H_2$ annihilates the state; $H_1+H_2$ instead evolves both copies upwards. Thus $O_1(t)=e^{iH_1t}O_1(0)e^{-iH_1t}$ and $O_2(t')=e^{-iH_2t'}O_2(0)e^{iH_2t'}$. CPT conjugation implies that second-copy  matrix elements satisfy $\,{}_2\langle\bar m|O_2(0)|\bar n\rangle_2=\langle n|O(0)|m\rangle$. The state sum then gives $O_2(t')|\mathrm{TFD}\rangle=O_1(t'+i\beta/2)|\mathrm{TFD}\rangle$. In addition,  $\langle\mathrm{TFD}|O_1(t)O_2(t')|\mathrm{TFD}\rangle=Z^{-1}\mathrm{Tr}[e^{-\beta H}O(t)O(t'+i\beta/2)]=Z^{-1}\mathrm{Tr}[e^{-\beta H}O(t-t'-i\beta/2)O(0)]$.  With $\tau=it$, the result is the Green's function $G(\beta/2+i(t-t'))$.}

Now we consider a radial null ray. In the two-sided geometry it is the ray of Figure~\ref{fig:penrose}. Sent in from a boundary, the ray reaches the singularity after a coordinate time
\begin{equation}
\mathrm{P}\!\!\int_0^\infty\frac{d\varrho}{f(\varrho)}=\frac\beta4\cot\frac\pi d .
\label{eq:nulltime}
\end{equation}
The integrand has a pole at the horizon, because the coordinate $t$ runs off to infinity there on both sides. The two infinities cancel, which is why the principal value is used. A ray that falls in, reflects off the singularity and goes to the second boundary therefore takes a real time $(\beta/2)\cot(\pi/d)$ altogether.

An imaginary part appears when we express this as a single-CFT thermal correlator: its Lorentzian separation is the real Killing-time difference minus $i\beta/2$. Taking that difference to be $(\beta/2)\cot(\pi/d)$ gives $t_c=(\beta/2)\cot(\pi/d)-i\beta/2$. In Euclidean time $\tau=it$ the separation is $\tc$, where
\begin{equation}
\tc=\frac\beta2+\frac{i\beta}{2}\cot\frac\pi d~.
\label{eq:bounce}
\end{equation}
Below, another time coordinate would be absolutely crucial:
$\tau_1=\tc-\beta$. In fact it will be more important than
$\tau_c$ (note that $\tau_1$ is  just a shift of $\tau_c$ by the thermal circle periodicity and it will take us to the second sheet).

Spacelike geodesics of large energy hug this null ray. As $t\to t_c$ their regularised length tends to $-\infty$, so $e^{-\Delta_O L}$ diverges, much as a correlator diverges on an ordinary light cone \cite{FHKS}. This is the bouncing singularity. It has been of interest because it is a sharp signature of the black-hole singularity in a well defined boundary observable. 

As we just mentioned, there is a crucial subtlety about where the singularity sits -- it does not dominate the two point function around Euclidean time $\tau_c$.  Indeed, $\tc$ lies in the middle of the strip $0<\Rea\tau<\beta$, where the physical correlator is analytic and bounded. 

We can call the strip $0<\Rea\tau<\beta$ the physical sheet. On the physical sheet, the light ray singularity actually cancels 
out. 
By periodicity in the Euclidean time, the physical sheet can be extended everywhere, but there are singularities along $\Rea\tau=0$. If we cross those singularities adiabatically then we end up at the second sheet with strip $-\beta<\Rea\tau<0$.

The bouncing geodesic contributes on this second sheet.  Start at positive Euclidean time and carry $\tau$ around the origin, across the real-time axis $\Rea\tau=0$, to the point $\tau_1=\tc-\beta$. The path has crossed the cut along the real-time axis, where the two operator orderings differ, and it arrives on another branch. The prediction of Einstein-Hilbert gravity is that on this branch the correlator has a singularity
\begin{equation}
G(\tau)\sim\frac{1}{(\tau-\tau_1)^{2\Delta_O}},\qquad
\tau_1=-\frac\beta2+\frac{i\beta}{2}\cot\frac\pi d,\qquad \Delta_O\gg1.
\label{eq:prediction}
\end{equation}
For $d=3,4,5$, $|\tau_1|<\beta$. The comparison of $|\tau_1|$ and $\beta$ will be the crux of the matter. (For $d=2$ there are various other issues.)

The singularity can also be seen without geodesics. In the thermal OPE the identity and the multi-stress-tensor operators contribute a power series in $(\tau/\beta)^d$, whose coefficients are fixed by the geometry \cite{FH,PSWambient,CLPV}.  On the physical sheet the double-trace operators cancel the singularity; indeed, as we said, on the physical sheet the correlator has to be bounded. On the analytically continued second sheet, though, there is no cancellation and, instead we see the singularity in the two point  function above. 

The subject has seen a great deal of recent activity. On the boundary side, Parisini, Skenderis and Withers~\cite[Sec.~5.4.3]{PSWambient} identified the large-order behavior of the stress-tensor sector of the thermal OPE and its relation to the bounce for $\Delta_O=3/2$ in $d=4$; this analysis was subsequently extended to other external dimensions in~\cite{CLPV}. For related work, see~\cite{FH,Huang,BGP1,BGP2,Valach,CLPV2,CV}. Its imprint on the full two- and four-point functions, on the thermal spectral function, and on line defects has been worked out \cite{AJCCM,Chak,JR1,JR2,AW1,GLS}. Related singularities have been tied to real and complex null geodesics that reflect off the singularity \cite{HLR,AH,DO,HLQZ,GVV,JK,AEJKP,AAL,GMV} and to quasinormal modes \cite{FL2,DIKZ,AW2,HZ,XNLKasner}. {Frenkel, Hartnoll, Kruthoff and Shi~\cite{FrenkelKasner} showed how Kasner exponents enter non-analytic corrections to thermal correlators through bouncing geodesics.} Asymptotic quasinormal modes have also been used to extract near-singularity scaling exponents and detect transitions between Kasner eons in semiclassical quantum black holes~\cite{CGHSSquantum,CGHSSinteriors}. The fate of the bounce away from the Einstein-Hilbert limit has been studied at finite coupling in the planar limit, through stringy corrections and the SYK model \cite{DIK,BCGHPR}. Related discussions include the absence of bulk-point singularities in exact four-point functions and the resolution at finite central charge of the ``forbidden singularities'' of semiclassical heavy-light Virasoro blocks~\cite{MSZ,FKLW}.\footnote{A complementary probe of black-hole interiors is the large-mass behavior of certain thermal one-point functions~\cite{GM}. Related information appears in interaction corrections to thermal two-point functions, as studied in~\cite{RGRsingularity,KRthermal}. Further developments include analytic studies of thermal one-point functions~\cite{BMonepoint,DK} and signatures of the proper time to the singularity in higher-point thermal correlators~\cite{Singhi}. For a de Sitter analogue  see~\cite{GKPdS}.}

What we do here, mainly, is to show that CFTs have an analyticity property on the second sheet. This not only rules out the bouncing singularity above at finite $G_N$, it also rules out simple-minded ways of smoothing this singularity, such as just shifting it around or making the singularity less severe.

We also put a rigorous bound on the Green's function of finite $N$ theories at the ``would be bounce.'' While there is no singularity in the boundary observable at finite $G_N$,  there could be a peak in the second sheet. We use two different methods to bound the would be singularity. One of the methods, we expect, is quite effective and yields a reasonable estimate that at finite $G_N$ the bounce singularity is bounded by an inverse power of $G_N$. 

Whether the bouncing singularity at finite $G_N$ is removed because the singularity {\it itself} is removed due to quantum gravity corrections or simply because the back-reaction of the infalling matter becomes very strong and the singularity is avoided, is hard to say from our arguments alone. It is also not clear if there is a real difference between the two scenarios, since if no observable can reach the singularity due to backreaction then one might as well say that there is no singularity.

\subsection{Setup}
We consider a unitary CFT with a unique normalised vacuum and a well-defined translation-invariant unique Gibbs state\footnote{Not every CFT has a unique Gibbs state in infinite volume. Thermal order can lead to spontaneous symmetry breaking at all nonzero temperatures and hence to multiple Gibbs states; see, e.g., \cite{ChaiThermalOrder,CCRthermal,CDSdiscrete,CDGSScontinuous,LRZfinite,HRSthermal,KPthermal,SYbiconical,CRSSbiconical}. For related discussions see \cite{BLSnonrestoration,BMSWinfinite,ANdeconfinement,BuchelThermal,BuchelFate,HHKLPentropic,HKLPSminimal,ANRSentropic,HKentropic}.}. Let $O$ be a Hermitian scalar primary of finite dimension $\Delta_O>0$ with a properly normalized two-point function. We denote \begin{equation}
G(\tau)=\vev{O(\tau,\mathbf0)O(0)}_\beta~.
\label{eq:setup}
\end{equation}
We will also frequently use $H(\tau)=\tau^{2\Delta_O}G(\tau)$ which has the property that $H(0)=1$. 
Real time is reached by setting $\tau=it$. The thermal manifold is $S^1_\beta\times\mathbb R^{d-1}$. (We consider the two operators at the same location in space, for simplicity.)
The physical correlator is analytic in the strip $0<\Rea\tau<\beta$, which we call the physical sheet. To see this, write $\tau=\tau_E+it$ with $\tau_E$ and $t$ real, so that $\tau_E$ is the Euclidean time and $t$ is the real time. Put the theory in a finite volume, where the spectrum is discrete. Let $|n\rangle$ be the eigenstates of the Hamiltonian $H$, with energies $\varepsilon_n$. Let $O_{mn}=\langle m|O(0)|n\rangle$ be the matrix elements of the operator. The operator at complex time is $O(\tau)=e^{\tau H}O(0)\,e^{-\tau H}$. Inserting a complete set of states  gives
\begin{equation}
G(\tau_E+it)=\frac1Z\sum_{m,n}e^{-(\beta-\tau_E)\varepsilon_m-\tau_E\varepsilon_n}
e^{it(\varepsilon_m-\varepsilon_n)}|O_{mn}|^2,
\qquad |G(\tau_E+it)|\leq G(\tau_E),\quad 0<\tau_E<\beta.
\label{eq:kms}
\end{equation}
Here $Z=\sum_ne^{-\beta\varepsilon_n}$ is the partition function. Every term is positive at $t=0$, and real time multiplies it by a phase. 

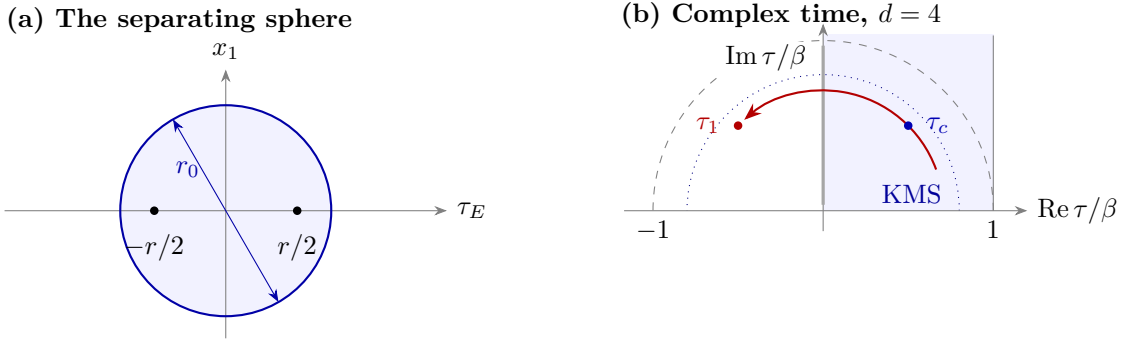
\begin{figure}[t]
\centering
\begin{tikzpicture}[>=Stealth,font=\small]
\begin{scope}[x=2.25cm,y=2.25cm]
\node[anchor=west,font=\bfseries\small] at (-1.35,1.12) {(a) The separating sphere};
\fill[blue!5] (0,0) circle (.62);
\draw[->,gray] (-1.3,0)--(1.3,0) node[right,black] {$\tau_E$};
\draw[->,gray] (0,-.75)--(0,.83) node[above,black] {$x_1$};
\draw[blue!65!black,thick] (0,0) circle (.62);
\fill (-.42,0) circle (.025) node[below=5pt] {$-r/2$};
\fill (.42,0) circle (.025) node[below=5pt] {$r/2$};
\draw[<->,blue!65!black] (-.31,.53694)--(.31,-.53694) node[pos=.4,above left=2pt] {$r_0$};
\end{scope}
\begin{scope}[shift={(7.9,0)},x=2.25cm,y=2.25cm]
\node[anchor=west,font=\bfseries\small] at (-1.25,1.15) {(b) Complex time, $d=4$};
\fill[blue!5] (0,0) rectangle (1,1.04);
\draw[dotted,blue!50!black] (.8,0) arc (0:180:.8);
\draw[dashed,gray] (1,0) arc (0:180:1);
\draw[->,gray] (-1.17,0)--(1.2,0) node[right,black] {$\Rea\tau/\beta$};
\draw[->,gray] (0,-.12)--(0,1.1);
\node[anchor=east,fill=white,inner sep=1pt] at (-.08,.90) {$\Ima\tau/\beta$};
\draw[very thick,gray!70] (0,.035)--(0,.97);
\draw[thin,gray] (1,0)--(1,1.03);
\draw[->,red!70!black,thick] (.664,.242) arc (20:131:.7071);
\fill[blue!70!black] (.5,.5) circle (.025) node[right=3pt] {$\tc$};
\fill[red!70!black] (-.5,.5) circle (.025) node[left=3pt] {$\tau_1$};
\node[below] at (1,0) {$1$};\node[below] at (-1,0) {$-1$};
\node[blue!60!black] at (.52,.10) {KMS};
\end{scope}
\end{tikzpicture}
\caption{(a) The insertions at $\tau_E=\pm r/2$ lie inside a flat ball of diameter $r_0$ (radius $r_0/2$), with $r<r_0<\beta$, embedded in the thermal cylinder. The operator pair creates a state on the sphere; its overlap with the thermal exterior gives the correlator. (b) The point $\tc$ is regular on the physical branch (shaded strip). We reach $\tau_1$ by analytic continuation to the second sheet.}
\label{fig:geometry}
\end{figure}

Hence we can immediately infer that the physical correlator is regular at $\tc$. We reach the point $\tau_1$ from positive Euclidean time by increasing $\arg\tau$ through the upper imaginary axis. In Figure~\ref{fig:geometry} we show how we reach the second sheet, e.g. for $d=4$, $\tau_1=\beta(-1+i)/2$. The  continuation can be taken entirely inside $0<|\tau|<\beta$, which will be very important to keep in mind. 

\subsection{Summary of the Argument}
The question is whether the singularity at $\tau_1$ survives at finite $N$.

At zero spatial separation the thermal OPE is a sum over scaling dimensions, $H(r)=\sum_\Delta a_\Delta (r/\beta)^\Delta$. It converges for real separations $0<r<\beta$ \cite{IKMPS,BBMMPthermal}. The reason is that for $r<\beta$ both insertions fit inside a flat ball, as in Figure~\ref{fig:geometry}(a), and radial quantisation in that ball gives a convergent expansion just as it does in the vacuum. 
Since the thermal OPE has essentially arbitrary nonnegative powers $\Delta$ which are not integer, to perform the analytic continuation we
write $\tau=re^{i\theta}$, with the understanding that the result is not $2\pi$ periodic. Term-by-term continuation gives
\begin{equation}
H(re^{i\theta})=\sum_\Delta a_\Delta\,(r/\beta)^\Delta\,e^{i\theta \Delta}.\label{phasesadded}
\end{equation}
We must show that this sum converges and gives an analytic function. 

This is not trivial at all: The coefficients $a_{\Delta}$ have no definite sign, so  convergence on the Euclidean time axis $\theta=0$ may rely on lots of cancellations. 
Multiplying terms by phases can actually make the sum divergent! (think of $\sum_n (-1)^n/n$, which becomes divergent if we multiply the odd terms by a minus sign).

The key is to compare with a slightly larger real separation $r_0$, with $r<r_0<\beta$. So the point is that the phases in~\eqref{phasesadded} can make the sum blow up a priori, but we know more! we know that it converges on the real axis for $r<r_0<\beta$ as well and such an $r_0$ always exists since $r<\beta$.

Let us analyze the partial sums at $r_0$ and take the supremum of their absolute values
\begin{equation}
M_{r_0}=\sup_{L}\Bigl|\sum_{0<\Delta\leq L}a_\Delta\,(r_0/\beta)^\Delta\Bigr|~.
\end{equation}
Convergence at $r_0$ ensures that $M_{r_0}$ is finite. Its $N$-dependence is hard to determine, since the partial sums can far exceed the full two-point function. We will say more about this below.

To reach our desired sum~\eqref{phasesadded} we need to just add on top of the convergent sum at $r_0$
the factors
\begin{equation}
(r/r_0)^\Delta\,e^{i\theta \Delta}=e^{-w\Delta},\qquad w=\log(r_0/r)-i\theta~.
\end{equation}
The relevant property of this weight is that it tends to zero exponentially fast at large $\Delta$. It is also important, to make a theorem out of it, that its derivative has a finite integral of its absolute value. Indeed, $|d e^{-wL}/dL|=|w|e^{-\log(r_0/r) L}$. Then we use the standard ``summation by parts trick'' to obtain
\begin{equation}
|H(re^{i\theta})-1|\leq M_{r_0}\int_0^\infty\Bigl|\frac{d}{dL}e^{-wL}\Bigr|\,dL
=M_{r_0}\sqrt{1+\theta^2/\log^2 (r_0/r)}.\label{boundsecondsheet}
\end{equation}
Geometrically, we obtain the length of the spiral traced by $e^{-wL}$ as it runs from $1$ towards $0$. Rotation certainly lengthens the path, but exponential decay keeps its length finite. We therefore established the existence of analytic continuation for $r<\beta$. This is applicable on the whole universal cover of the physical sheet for $0<|\tau|<\beta$, not just the second sheet.

Section~\ref{sec:main} reviews this little trick concerning partial sums. 

The bounce sits at $|\tau_1|=\beta/(2\sin(\pi/d))$, which in $d=3,4,5$ dimensions can be reached inside the disc of radius $\beta$, as in Figure~\ref{fig:geometry}(b), hence there is no singularity at $\tau_1$ at finite $N$. 

A rough estimate suggests that with the best choice of $r_0$ the bound~\eqref{boundsecondsheet} on the correlator at $\tau_1$ grows as a power of $1/G_N$. 

\paragraph{How is this compatible with Einstein-Hilbert gravity?}
In the gravity approximation the correlator is the sum of two series, $G=G_T+G_{[OO]}$. The first collects the identity and the multi-stress tensors, $G_T=\tau^{-2\Delta_O}\sum_n\Lambda_n(\tau/\beta)^{dn}$. The second collects the double traces and is a series in even integer powers of $\tau$. In $d=4$ we have $\Lambda_n\sim(-4)^nn^{2\Delta_O-3}$, so the series has radius of convergence $4^{-1/4}\beta=\beta/\sqrt2=|\tc|$, and $G_T$ is singular at $\tc$. The full $G$ is regular at $\tc$ on the physical branch. So $G_{[OO]}$ must carry the opposite singularity, and its series has the same radius of convergence. 

Going round the origin to the second branch multiplies $G_T$ by $e^{-2\pi i\Delta_O}$ and leaves the integer powers in $G_{[OO]}$ untouched. The cancellation is now spoiled and we have a singularity at $\tau_1$.

How can this be compatible with our bound~\eqref{boundsecondsheet}?
We required convergence at some real $r_0>|\tau_c|$,
as follows from locality of CFT. 
The pure gravity theory is not compatible with this requirement. 
This is a failure of the gravity approximation  already at real Euclidean time. 

The Green's function $G=G_T +G_{[OO]}$ exists on the physical sheet. The issue is that its expansion around $\tau=0$ along the Euclidean time  does not have the required radius of convergence.
For instance, if we order the terms from $G_T +G_{[OO]}$ by scaling dimension (choosing $\Delta_O$ as to avoid degeneracies with the energy-momentum tensor), there would be terms in $G_T$ that scale as $(\sqrt 2 r/\beta)^{4n}$ and hence the partial sums  grow indefinitely for $r>\beta/\sqrt2$, which is why the OPE fails to converge, even if there is no singularity on the physical sheet.  

At finite $N$ (and perhaps also at finite string tension, in some cases) we must recover complete OPE convergence out to $\beta$. So the exact OPE data, meaning the full list of dimensions and coefficients, must depart from the Einstein-Hilbert calculation at sufficiently high dimension. 
The departure is small but not uniformly small. For instance, at finite $N$ the many operators built from the stress tensors surely acquire different anomalous dimensions so the decomposition $G_T+G_{[OO]}$ loses its meaning. 

So the problem with the gravity approximation manifests itself well before we get to discuss the bouncing singularity on the second sheet. Already on the first sheet, the gravity approximation does not lead to the correct radius of convergence of the OPE and the two issues are deeply connected. 

\paragraph{More Bounds}
We already explained why the finite $N$  correlator is finite at $\tau_1$. We bounded it by the partial sums~\eqref{boundsecondsheet}.  There is another bound, which is perhaps more interesting but at the same time in practice is probably very weak.

We cut the thermal path integral along a sphere of diameter $r_0$ that encloses the two operators. The ball with the two insertions prepares a state $\ket{u(r)}$ on the sphere, the rest of the thermal cylinder prepares a state $\langle\Psi_{r_0}|$, and the correlator is their overlap. Analytical continuation inserts the unitary operator $e^{i\theta D}$ between them, with $D$ standing for dilatations. The overlap of two normalisable states is bounded by the product of their norms, so \begin{equation}
|H(re^{i\theta})-1|\leq\sqrt{[F-1]\,[X-1]},\qquad F=\|\Psi_{r_0}\|^2,\quad X=\|u(r)\|^2 .
\label{eq:intro-bound}
\end{equation}
Actually the norm of $\ket{u(r)}$
is just the vacuum four-point function, which is finite. The norm $F$ is the sum of the squares of all thermal one-point functions. Equivalently, it is a ratio of partition functions on a connected sum manifold
\begin{equation}
F=e^{\mathcal A}\,\frac{Z[\mathcal M_{r_0}]}{Z[M]^2}.
\label{eq:intro-F}
\end{equation}
Here $M$ is the thermal cylinder. The manifold $\mathcal M_{r_0}$ (not to be confused with the partial sums $M_{r_0}$) is made from two copies of $M$: remove a ball of diameter $r_0$ from each and join them along the two spheres via a connected sum, as in Figure~\ref{fig:states}(c).  It has two independent thermal circles and the sphere at the neck is non-contractible. The factor $e^{\mathcal A}$ represents local counter-terms. The extensive free energies cancel in the ratio. For finite $N$ theories such partition functions must be finite up to local counter-terms. {The same connected-sum geometry and its expansion in squared thermal one-point coefficients were studied by Simmons-Duffin and Xu~\cite{SDXgenusTwo} in the thermal flat limit of their genus-two crossing equation.} 

$X$ is just a vacuum four point function and approaches that of generalized free field theory. By contrast, we expect that the partition function $F$ would roughly scale as the exponent of the heat capacity $\log F\sim f h(r_0/\beta)$ with $h>0$. The heat capacity per unit volume grows as $f\sim N^2$ in many holographic theories, hence the divergence in $F$ at infinite $N$. This exponential bound on the bouncing geodesic is perhaps neat, but practically very loose, we expect.\footnote{Amusingly, the divergence of $F$ at infinite $N$ is not sufficient to obtain bouncing singularities. In the critical $O(N)$ model $F$ also diverges with $N$, yet there is (most likely) no bounce and the correlator is finite on the second sheet.}

I was informed that Simon Caron-Huot, David Simmons-Duffin and Douglas Stanford have independently obtained related results~\cite{CHSprivate}.
In addition, shortly after this paper appeared,  Batra, Levine and Shenker~\cite{BLSbackreaction} showed in two-dimensional dilaton gravity that gravitational backreaction smooths out bouncing-geodesic singularities.

\section{Thermal OPE and Analytic Continuation}\label{sec:main}

\subsection{Euclidean convergence}\label{sec:physics}
The standard Euclidean thermal-OPE result of \cite{IKMPS}: the expansion converges when the insertions lie inside a flat ball embedded in $S^1_\beta\times\mathbb R^{d-1}$. For separation along the thermal circle this gives
\begin{equation}
0<r<\beta.
\label{eq:euclidean-domain}
\end{equation}

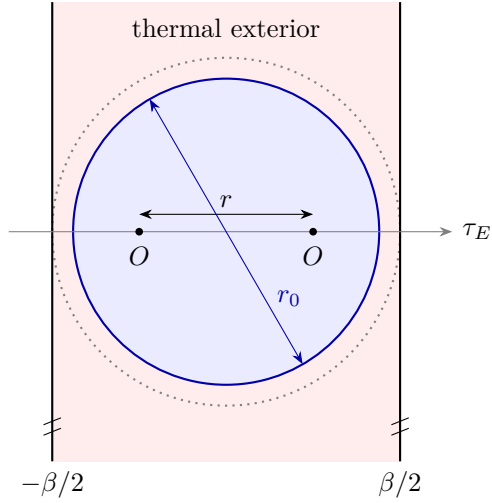
\begin{figure}[!ht]
\centering
\begin{tikzpicture}[>=Stealth,font=\small,x=2.3cm,y=2.3cm]
\fill[red!7] (-1,-1.32) rectangle (1,1.32);
\fill[white] (0,0) circle (.88);
\fill[blue!8] (0,0) circle (.88);
\draw[dotted,thick,gray] (0,0) circle (1);
\draw[blue!65!black,thick] (0,0) circle (.88);
\draw[thick] (-1,-1.32)--(-1,1.32);
\draw[thick] (1,-1.32)--(1,1.32);
\draw (-1.05,-1.17)--(-.95,-1.13);\draw (-1.05,-1.11)--(-.95,-1.07);
\draw (.95,-1.17)--(1.05,-1.13);\draw (.95,-1.11)--(1.05,-1.07);
\draw[->,gray] (-1.25,0)--(1.3,0) node[right,black] {$\tau_E$};
\fill (-.5,0) circle (.022) node[below=2pt] {$O$};
\fill (.5,0) circle (.022) node[below=2pt] {$O$};
\draw[<->] (-.5,.1)--(.5,.1) node[pos=.5,above=-1pt] {$r$};
\draw[<->,blue!65!black] (-.44,.762)--(.44,-.762) node[pos=.75,right=1pt] {$r_0$};
\node at (0,1.17) {thermal exterior};
\node at (-1,-1.45) {$-\beta/2$};
\node at (1,-1.45) {$\beta/2$};
\end{tikzpicture}
\caption{The geometry behind \eqref{eq:euclidean-domain}. The strip is the thermal cylinder, with the two vertical edges identified. The operators sit on the thermal circle, a distance $r$ apart. A ball about their midpoint is flat until it meets its own image, at radius $\beta/2$ (dotted). For $r<\beta$ one can choose $r<r_0<\beta$. The ball of diameter $r_0$ contains the operators, while the region outside it supplies a certain interesting  thermal state, whose overlap with the state produced by the insertions gives our desired Green's function. }
\label{fig:shell}
\end{figure}

Figure~\ref{fig:shell} shows our setup -- the auxiliary diameter $r_0$, with $r<r_0<\beta$, appeared prominently in the introduction and will soon appear again.  

Translation invariance makes the one-point functions of descendants vanish. At zero spatial separation the thermal OPE is therefore, as in the introduction
\begin{equation}
H(r)=\sum_{\Delta}a_\Delta(r/\beta)^\Delta~,
\label{eq:thermalOPE}
\end{equation}
where $a_0=1$ and $H$ is related to the Green's function by a factor of $r^{2\Delta_O}$. 
As usual, the $\Delta$ are non-negative with only finitely many distinct $\Delta$ below any cutoff. The coefficients $a_\Delta$ need not be positive. Our challenge is to understand what happens as we analytically continue this sum $\tau=re^{i\theta}$. 
We will not assume absolute convergence of~\eqref{eq:thermalOPE}; the argument below uses only convergence of the dimension-ordered sum.

What we will do is to use the fact that the sum also converges at Euclidean time with $r<r_0<\beta$. 

The argument is the standard convergence theorem for general Dirichlet series, proved using   summation by parts.\footnote{One can read more about this in \cite[Ch.~II, Sections~1--2]{HR}. We thank Claude Code for finding this very illuminating source and for checking that our argument is indeed mathematically defensible.} Define the partial sums at the auxiliary separation $r_0$ (identity omitted)
\begin{equation}
A_{r_0}(L)=\sum_{0<\Delta\leq L}a_{\Delta}\,(r_0/\beta)^{\Delta},
\qquad M_{r_0}=\sup_{L}|A_{r_0}(L)|.
\label{eq:boundedpartials}
\end{equation}
Convergence at $r_0$ implies $M_{r_0}<\infty$. Denote
\begin{equation}
w=\log(r_0/r)-i\theta,\qquad \log(r_0/r)>0.
\label{eq:w}
\end{equation}
Then we can rewrite $H(re^{i\theta})$ in the following way  $1+\sum_{\Delta>0}a_{\Delta}\,(r_0/\beta)^{\Delta}\,e^{-w\Delta}$.
What this shows is that our analytically continued sum can be obtained by starting from a convergent sum and then multiplying each term by $e^{-w\Delta}$. 

We still have to prove that if we start from a convergent sum and then multiply each term by $e^{-w\Delta}$ then we still get a convergent sum. This is again not obvious due to the oscillations. 

We appeal to the summation by parts trick which  gives the identity
\begin{equation}
\sum_{0<\Delta\leq L}a_{\Delta}\,(r_0/\beta)^{\Delta}\,e^{-w\Delta}
=A_{r_0}(L)e^{-wL}+w\int_0^L A_{r_0}(x)e^{-wx}\,dx.
\label{eq:abel-finite}
\end{equation}
The partial sums used here are those of the Euclidean OPE at separation $r_0$, with the identity term omitted. Order the positive dimensions as $0<\Delta_1<\Delta_2<\cdots$. By definition, $|A_{r_0}(\Delta_j)|\leq M_{r_0}$.
Substituting $a_{\Delta_j}\,(r_0/\beta)^{\Delta_j}=A_{r_0}(\Delta_j)-A_{r_0}(\Delta_{j-1})$, with $\Delta_0=0$ and $A_{r_0}(0)=0$, into the finite sum makes it telescope. Replacing the resulting differences by integrals of the derivative of $e^{-wx}$ gives \eqref{eq:abel-finite}. The function $A_{r_0}(x)$ is constant between neighboring dimensions. 

The boundary term obeys $|A_{r_0}(L)e^{-wL}|\leq M_{r_0}e^{-L\log(r_0/r)}$ and vanishes as $L\to\infty$. The remaining integral converges absolutely, since
\begin{equation}
\int_0^\infty|wA_{r_0}(x)e^{-wx}|\,dx
\leq M_{r_0}|w|\int_0^\infty e^{-x\log(r_0/r)}\,dx
=M_{r_0}\frac{|w|}{\log(r_0/r)}.
\label{eq:weight-integral}
\end{equation}
The Green's function  therefore exists and satisfies the following identity for every auxiliary separation $r_0$ with $r<r_0<\beta$:
\begin{equation}
 H(\tau=r e^{i\theta})-1=w\int_0^\infty A_{r_0}(x)e^{-wx}\,dx,
\qquad |H(\tau)-1|\leq M_{r_0}\frac{|w|}{\log(r_0/r)}.\quad
\label{eq:abel}
\end{equation}
Note that the ``total length'' $\int_0^\infty |w|e^{-x\log(r_0/r)}\,dx=|w|/\log(r_0/r)$ appears prominently. When we start from a convergent sum and multiply it by terms that decay and have phases, it is important that this ``total length'' converges for the resulting sum to converge.\footnote{For example, $b_n=(-1)^n$ has bounded partial sums and $f(x)=e^{i\pi x}/(1+x)$ is smooth and tends to zero, but $\sum_{n\geq1}b_nf(n)=\sum_{n\geq1}(1+n)^{-1}$ diverges. Here the ``total length'' $\int_0^\infty|f'(x)|\,dx$ is infinite.} 

We can also see how fast the OPE converges on the second sheet.
Subtracting \eqref{eq:abel-finite} from \eqref{eq:abel} gives
\begin{equation}
\left|H(r e^{i\theta})-\sum_{\Delta\leq L}a_{\Delta}\,(r_0/\beta)^{\Delta}\,e^{-w\Delta}\right|
\leq M_{r_0}e^{-L\log(r_0/r)}\left(1+\frac{|w|}{\log(r_0/r)}\right).
\label{eq:abel-error}
\end{equation}
(Note that now the unit operator is included in the left-hand side.)
The right-hand side then tends to zero as $L\to\infty$ which guarantees convergence and analyticity on the second sheet, as long as $r<\beta$.

We have therefore proved analyticity on the second sheet up to $|\tau|<\beta$ and in particular excluded the bounce singularity in $d=3,4,5$. (For related Tauberian estimates of thermal OPE asymptotics, truncation errors and Euclidean two-point functions, see~\cite{MMPtauberian}.)

The $N$-dependence of the bounds \eqref{eq:abel}--\eqref{eq:abel-error} is hard to determine: they depend on the largest OPE partial sum, which can far exceed the full two-point function. It is  nevertheless important to estimate how the bound scales with $G_N$. Take $d=4$, where the stress-tensor coefficients of the gravity approximation are known, $\Lambda_n\sim(-4)^nn^{2\Delta_O-3}$, see \eqref{eq:large-order} below. At a real separation $r_0>|\tc|=\beta/\sqrt2$ the terms $\Lambda_n(r_0/\beta)^{4n}$ have magnitude $n^{2\Delta_O-3}(r_0/|\tc|)^{4n}$. They alternate in sign and they grow, so a partial sum is of the size of its last term. This is why $M_{r_0}$ is infinite at $G_N=0$ and why the classical gravity approximation fails already on the physical sheet. At finite $G_N$ the growth has to stop (or be canceled, somehow). Let $n_*$ be the order at which, crudely speaking, the exact coefficients depart from the gravity values $\Lambda_n$ such that the estimates of the partial sums deviate from the gravity approximation.
Then $M_{r_0}\sim n_*^{2\Delta_O-3}(r_0/|\tc|)^{4n_*}$. 

Now use \eqref{eq:abel} at $\tau=\tau_1$, where $r=|\tau_1|=|\tc|$ and $\theta=3\pi/4$. Taking $r_0$ just slightly bigger than $|\tc|$, the bound reads $|H(\tau_1)-1|\lesssim n_*^{2\Delta_O-3}\,e^{4n_*\log(r_0/r)}\,\theta/\log(r_0/r)$. For fixed $r_0$ it is exponential in $n_*$, but we are free to choose $r_0$. The best choice is $\log(r_0/r)=1/(4n_*)$, that is, $r_0$ only slightly larger than $|\tau_1|$, and it gives $|H(\tau_1)|\lesssim n_*^{2\Delta_O-2}$ up to a constant.  In units of the AdS radius $C_T\propto1/G_N$. Now let us argue that $n_*\sim (C_T)^{\#}$, with $\#>0$ some positive power. Consider operators made of $n$ stress tensors -- we expect corrections of relative size $n^2/C_T$, as due to pairwise interactions among $n$ gravitons. Then $n_*~\sim C_T^{1/2}$. This is extremely crude, but any other similar set of considerations show that $n_*$ has to be some inverse power of the Newton constant. So we expect something like  $|H(\tau_1)|\lesssim (G_N)^{-2\# \Delta_O}$.   Another very crude idea would be that perhaps when the curvature is Planckian, the geodesic magically turns back. A geodesic of energy $E$ then turns around at $\varrho=\varrho_h^2/E$, the curvature there is Planckian when $\beta E\sim C_T^{1/6}$. For $\tau=\tau_1(1-\epsilon)$ with $0<\epsilon\ll1$, the gravity OPE terms behave as $n^{2\Delta_O-3}e^{-4n\epsilon}$, so at fixed $\Delta_O$ the relevant orders are $n\sim\epsilon^{-1}\sim\beta/|\tau-\tau_1|$. The large-energy bouncing geodesic obeys $|\tau-\tau_1|\sim E^{-1}$~\cite{CLPV}. {See also~\cite[Eq.~(19)]{FrenkelKasner} for this large-energy relation in flows to Kasner interiors.}  Matching these scales at the Planckian cutoff suggests $n_*\sim\beta E\sim C_T^{1/6}$. Under this crude assumption we would obtain $
|H(\tau_1)|\lesssim  G_N^{-\Delta_O/3}$. 

In summary, our bound using partial sums seems effective in the sense that very crude models for how the singularity is resolved  lead to sensible bounds on the Green's function on the second sheet.

\section{Relation to the Leading Gravity Approximation}\label{sec:planar}

In the leading gravity description of a scalar probe the thermal correlator splits as
\begin{equation}
G=G_T+G_{[OO]},\qquad
G_T(\tau)=\tau^{-2\Delta_O}\sum_{n\geq0}\Lambda_n(\tau/\beta)^{dn},
\qquad G_{[OO]}(\tau)=\sum_{k\geq0}c_k\tau^{2k}.
\label{eq:split}
\end{equation}
The first piece collects the identity and multi-stress tensors, and the second double traces. 

We take $\Delta_O$ away from  resonances,
$dn\neq 2\Delta_O+2k$ so the sectors can be distinguished by their exponents. This does not affect the conclusions and makes the discussion easier to follow.

For an analytic thermal-bootstrap approach to holographic correlators that combines the KMS condition with multi-stress-tensor OPE data and treats integer external dimensions, see~\cite{BBMMPholography}.

{For generic non-integer $\Delta_O$ in four dimensions, the high-order calculations and numerical asymptotic analysis of~\cite{CLPV}, extending the $\Delta_O=3/2$ analysis of~\cite{PSWambient}, give the behavior}
\begin{equation}
\Lambda_n\sim C(\Delta_O)\,(-4)^n n^{2\Delta_O-3}
\qquad\Longrightarrow\qquad G_T(\tau)\sim\frac{C(\Delta_O)}{(\tc-\tau)^{2\Delta_O-2}}.
\label{eq:large-order}
\end{equation}
Here $C(\Delta_O)$ is the dimensionless coefficient. For $C(\Delta_O)\neq0$, the series has radius $\beta/\sqrt2=|\tc|$. The full correlator is regular on the first sheet and in particular at $\tc$ by \eqref{eq:kms}, so the double-trace contribution must cancel the stress-sector singularity.

Starting near the positive real axis and continuing through the upper half-plane, we dial the phase of $\tau$  through $\pi/2$ to $\pi/2<\arg\tau<\pi$. 

The continued correlator, $\Gc(\tau)$, can be compared with $G(-\tau)$ which still lies in the physical sheet for 
$\Rea\tau<0$
\begin{equation}
\Gc(\tau)-G(-\tau)
=\sum_{\Delta} a_{\Delta}\,\beta^{-\Delta}(-\tau)^{p_{\Delta}}\bigl(e^{i\pi p_{\Delta}}-1\bigr),
\qquad p_{\Delta}=\Delta-2\Delta_O.
\label{eq:disc}
\end{equation}
Due to the exponents in \eqref{eq:split}, the double traces have $p=2k$ and acquire no phase. The stress tensor contributions have $p=dn-2\Delta_O$; and so, for even $d$, every term acquires the phase $e^{-2\pi i\Delta_O}$. Consequently, \begin{equation}
\quad \Gc(\tau)=G(-\tau)
+\bigl(e^{-2\pi i\Delta_O}-1\bigr)\,G_T(-\tau)~.
\label{eq:monodromy}
\end{equation}
While the singularities of $G_T$ and $G_{[OO]}$ cancel in the physical correlator $G(-\tau)$, the relative phase in $\Gc(\tau)$ changes this cancellation, leaving the stress-sector singularity multiplied by $e^{-2\pi i\Delta_O}-1$. 

We see that the leading order approximation to gravity violates the allowed behavior in  finite $N$ CFT. The violation is already on the Euclidean time axis, as the radius of convergence is not right. The issue is that taking large $N$ first does not commute with the OPE.  In other words, at any finite $C_T$ the exact data must depart from the gravity data at sufficiently large dimension and the limits do not commute.

One disadvantage of our analysis thus far is that the bounds we placed on the Green's function on the second sheet depend on an auxiliary $r_0$ and also they depend on the angle $\theta$. 
Next we will discuss a new bound that has no $\theta$ dependence.

\section{Bounds from Partition Functions}\label{sec:quantitative}

The constant $M_{r_0}$ which appeared in our bound~\eqref{eq:abel} is  unknown, and it depends on an auxiliary $r_0$, and also the bound \eqref{eq:abel} grows with the  angle $\theta$. 

By contrast, the bound below is $\theta$ independent. But the price to pay is that, since it is basically a Cauchy--Schwarz type inequality,  it misses out on all the cancellations and we expect it to be very suboptimal.  Therefore we expect this bound would be very weak, but it is still worth presenting it. 

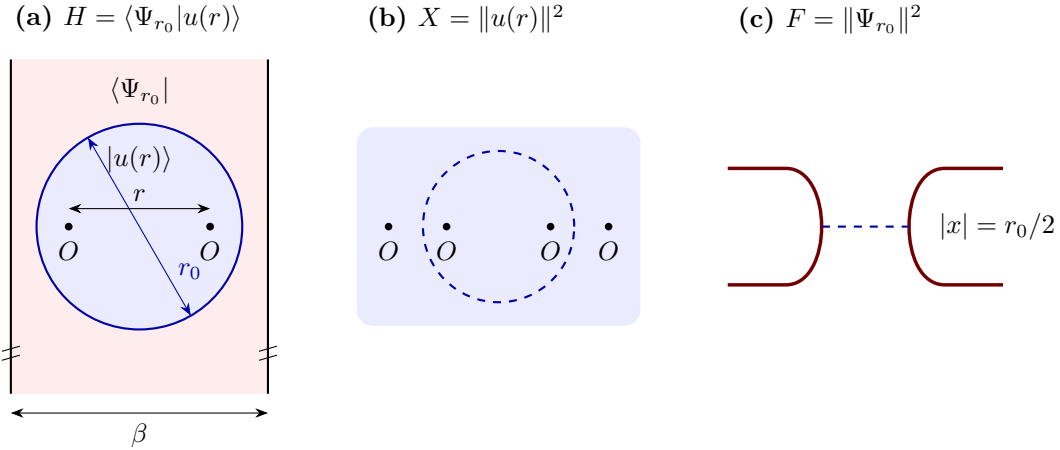
\begin{figure}[!ht]
\centering
\begin{tikzpicture}[>=Stealth,font=\small]
\begin{scope}[x=1.7cm,y=1.7cm]
\node[anchor=west,font=\bfseries\small] at (-1.05,1.62) {(a) $H=\langle\Psi_{r_0}|u(r)\rangle$};
\fill[red!7] (-1,-1.3) rectangle (1,1.3);
\fill[white] (0,0) circle (.8);
\fill[blue!8] (0,0) circle (.8);
\draw[blue!65!black,thick] (0,0) circle (.8);
\draw[thick] (-1,-1.3)--(-1,1.3);
\draw[thick] (1,-1.3)--(1,1.3);
\draw (-1.07,-1.04)--(-.93,-1.00);\draw (-1.07,-.97)--(-.93,-.93);
\draw (.93,-1.04)--(1.07,-1.00);\draw (.93,-.97)--(1.07,-.93);
\fill (-.55,0) circle (.03) node[below=2pt] {$O$};
\fill (.55,0) circle (.03) node[below=2pt] {$O$};
\draw[<->] (-.55,.14)--(.55,.14) node[midway,above=-1pt] {$r$};
\draw[<->,blue!65!black] (-.4,.69282)--(.4,-.69282) node[pos=.75,right=1pt] {$r_0$};
\node at (0,.52) {$|u(r)\rangle$};
\node at (0,1.06) {$\langle\Psi_{r_0}|$};
\draw[<->] (-1,-1.45)--(1,-1.45) node[midway,below] {$\beta$};
\end{scope}
\begin{scope}[shift={(4.75,0)},x=1.25cm,y=1.25cm]
\node[anchor=west,font=\bfseries\small] at (-1.5,2.2) {(b) $X=\|u(r)\|^2$};
\fill[blue!8,rounded corners=6pt] (-1.5,-1.05) rectangle (1.5,1.05);
\draw[blue!65!black,thick,dashed] (0,0) circle (.8);
\fill (-.55,0) circle (.04) node[below=2pt] {$O$};
\fill (.55,0) circle (.04) node[below=2pt] {$O$};
\fill (-1.164,0) circle (.04) node[below=2pt] {$O$};
\fill (1.164,0) circle (.04) node[below=2pt] {$O$};
\end{scope}
\begin{scope}[shift={(9.6,0)},x=1.4cm,y=1.4cm]
\node[anchor=west,font=\bfseries\small] at (-1.3,1.96) {(c) $F=\|\Psi_{r_0}\|^2$};
\draw[red!45!black,line width=1.3pt] (-1.3,.55)--(-.75,.55) .. controls (-.3,.55) and (-.3,-.55) .. (-.75,-.55)--(-1.3,-.55);
\draw[red!45!black,line width=1.3pt] (1.3,.55)--(.75,.55) .. controls (.3,.55) and (.3,-.55) .. (.75,-.55)--(1.3,-.55);
\draw[blue!65!black,thick,dashed] (-.41,0)--(.41,0);
\node[anchor=west] at (.6,0) {$|x|=r_0/2$};
\end{scope}
\end{tikzpicture}
\caption{(a) The thermal cylinder, with its two vertical edges identified, cut along a sphere of diameter $r_0$ around the midpoint of the operators. The ball with the two insertions prepares the state $|u(r)\rangle$ on the sphere. The rest of the cylinder, which has no insertions, prepares $\langle\Psi_{r_0}|$. Gluing them back gives the correlator. (b) The norm of $|u(r)\rangle$: the ball glued to its reflection through the sphere is flat space with four insertions. (c) The norm of $\Psi_{r_0}$: the thermal exterior glued to its reflection. This is two copies of the thermal cylinder joined through a neck, drawn here in cross-section.}
\label{fig:states}
\end{figure}

We now follow the construction of Figure~\ref{fig:states}. Draw a sphere of diameter $r_0$  around the midpoint of the two operators. Take it large enough to enclose them and small enough not to meet its own image, $r<r_0<\beta$, and cut the path integral along it, as in Figure~\ref{fig:states}(a). Each piece prepares a state of the fields on the sphere. The ball, with the two insertions, prepares the state $\ket{u(r)}$ -- it is the exact same state that the operator pair creates from the vacuum.  The rest of the thermal cylinder has no insertions, and prepares a second state $\langle\Psi_{r_0}|$.  On the thermal cylinder it is the vacuum plus excitations, such that its component along an operator $\mathcal O$ is the thermal one-point function $\vev{\mathcal O}_\beta$. Gluing the two pieces back together is the overlap $H(r)\sim\langle\Psi_{r_0}|u(r)\rangle$ which is related to our Green's function.

Changing $r$ to $re^{i\theta}$ inserts $e^{i\theta D}$ between the two states. This $e^{i\theta D}$ acts on the state $|u(r)\rangle$ in the same way as in the vacuum. This operator is unitary.  The overlap of two states is bounded by the product of their norms. So if both states are normalisable, we obtain a bound that does not depend on $\theta$.
The two norms are themselves path integrals, shown in Figure~\ref{fig:states}(b,c). To square a state one glues its path integral to the reflected copy. For $\ket{u(r)}$ this gives flat space with four insertions, so its norm  is a vacuum four-point function. It is finite. For $\Psi_{r_0}$ it gives two copies of the thermal exterior joined through a sphere. Its norm $F$ is the sum of the squares of all thermal one-point functions. As a path integral it is the ratio \eqref{eq:intro-F} of the partition function on the connected sum manifold to the square of the thermal partition function. That $F$ is finite should be regarded as general fact about QFT at finite $N$ -- which up to counter-terms has finite partition functions. This is essentially an axiom that holds in well behaved theories, see e.g.~\cite{McNamaraWang} and the Kontsevich-Segal axioms. (Our manifold is non-compact, but we are normalizing it by manifolds with the same non-compact tails and therefore, by locality, it must be well behaved. We could also make space compact, if we wanted.)

Let $r<r_0<\beta$. The exterior state is normalised by
\begin{equation}
F(s):=\|\Psi_{r_0}\|^2<\infty,\qquad
s=\frac{r_0}{\beta},\qquad \langle\Psi_{r_0}|0\rangle=1.
\label{eq:Fdefinition}
\end{equation}
The interior state is
\begin{equation}
\ket{u(r)}=r^{2\Delta_O}O(r/2)O(-r/2)\ket0,
\qquad \langle0|u(r)\rangle=1,
\qquad H(r)=\langle\Psi_{r_0}|u(r)\rangle .
\label{eq:u}
\end{equation}
The norm is an ordinary vacuum four point function. Denote $\rho=(r/r_0)^2$, then
\begin{equation}
X(\rho):=\|u(r)\|^2=\sum_{\Delta}q_{\Delta}\,\rho^{\Delta}.
\label{eq:Xdef}
\end{equation}
$q_\Delta\geq0$ and $q_0=1$. This vacuum four-point function is finite for $r<r_0$~\cite{PRER}.\footnote{For a generalised free field, we get $1+[4\rho/(1+\rho)^2]^{2\Delta_O}+[4\rho/(1-\rho)^2]^{2\Delta_O}$. The last term diverges when the insertions meet their reflections, at $\rho=1$.}

Analytic continuation acts on the interior via $e^{i\theta D}$ while the thermal exterior remains fixed:
\begin{equation}
H(re^{i\theta})=\langle\Psi_{r_0}|e^{i\theta D}|u(r)\rangle
=\sum_\Delta a_\Delta(r/\beta)^\Delta e^{i\theta\Delta}.
\label{eq:H}
\end{equation}
Both states have unit vacuum component, which we can subtract, so \[
\bigl\|\ket{\Psi_{r_0}}-\ket0\bigr\|^2=F(s)-1,
\qquad
\bigl\|\ket{u(r)}-\ket0\bigr\|^2=X(\rho)-1.
\]
Since $e^{i\theta D}$ is unitary and leaves $\ket0$ fixed, Cauchy--Schwarz gives
\begin{equation}
\begin{aligned}
|H(re^{i\theta})-1|\leq \sqrt{[F(s)-1]\,[X(\rho)-1]},\qquad r<r_0.
\end{aligned}
\label{eq:bound}
\end{equation}
This holds for every real $\theta$, but, as mentioned above, we do not necessarily expect this to be a tight or even parametrically correct bound.
From general principles, we expect $\log F\sim N^2$ in holographic theories, so this gives an exponentially large bound at weak Newton constant. The bound using partial sums, by contrast, is expected to be an inverse power of the Newton constant.

{Simmons-Duffin and Xu~\cite[Sec.~3.1, Eqs.~(51)--(52)]{SDXgenusTwo} derived the expansion below in the thermal flat limit of their genus-two partition function; in their notation, $F(s)=h(s^2/4)$.  Here we use this partition function to bound the analytically continued thermal two-point function.}

The partition function $F(s)$ is just the sum of squares of one-point functions in the thermal state, concretely, for unit normalized operators, $\vev{\mathcal O^{\mu_1\ldots\mu_J}}_\beta
=b_{\mathcal O}\beta^{-\Delta_{\mathcal O}}
(e^{\mu_1}\!\cdots e^{\mu_J}-\text{traces})$,
where $e^\mu$ is the unit vector along the thermal circle. For $d>2$, we get
\begin{equation}
F(s)=1+\sum_{\mathcal O\ne\mathbf1}
|b_{\mathcal O}|^2\kappa_J\,(s/2)^{2\Delta_{\mathcal O}},
\qquad \kappa_J=\frac{(d-2)_J}{2^J(d/2-1)_J} ,
\label{eq:F}
\end{equation}
Here $(a)_J=a(a+1)\cdots(a+J-1)$ with $(a)_0=1$. This is a positive sum over all thermal one-point functions. It converges at finite $N$.  (One term is known exactly in terms of the heat capacity per unit volume,  due to the energy momentum tensor, but it is a tiny contribution of order $N^2$ while we expect the whole sum to be exponential of $N^2$.) $\|\Psi\|^2$
really captures all the thermal states while the probe could see only a small fraction of them, so that is why one should not expect the upper bound by $F$ to be useful practically.
The critical $O(N)$ model provides another indication that this bound is very loose. We do not expect a bounce in the $O(N)$ model but the upper bound is still an exponent of $N$.

\section*{Acknowledgments}
I am very grateful to Ahmed Almheiri, Alexander Frenkel, and Alessio Miscioscia for a series of lectures on bouncing geodesics, where I learned about this problem. I am also grateful to the three of them for making incisive comments on the draft. I gratefully acknowledge NSF Award Number 2310283. I used Claude Code and ChatGPT for some editing, figures, and for verifying the rigor of the claims.

\end{document}